\documentclass[aps,prl,10pt,twocolumn,floats,letterpaper,showpacs]{revtex4-1}

\usepackage[T1]{fontenc}
\usepackage{graphicx}
\usepackage{mathrsfs}
\usepackage[intlimits,centertags]{amsmath}
\usepackage{amssymb,amsfonts}
\usepackage[pdftex]{hyperref}
\usepackage[x11names]{xcolor}
\usepackage{aas_macros}

\newcommand{\bbeta}{{\boldsymbol \beta}}

\hypersetup{pdftitle={Deciphering the Dipole Anisotropy of Galactic Cosmic Rays},
pdfsubject={Deciphering the Dipole Anisotropy of Galactic Cosmic Rays},
pdfauthor={Markus Ahlers},
pdfstartview={FitH},
colorlinks=true,
bookmarksopen=false,
bookmarksnumbered=false,
bookmarksopenlevel=0,
linkcolor=Blue1!60!black,
citecolor=Green1!50!black,
urlcolor=Blue1!70!black
}

\begin{document}

\title{Deciphering the Dipole Anisotropy of Galactic Cosmic Rays}
\author{Markus Ahlers}
\affiliation{WIPAC \& Department of Physics, University of Wisconsin--Madison, Madison, WI 53706, USA}
\begin{abstract} 
Recent measurements of the dipole anisotropy in the arrival directions of Galactic cosmic rays (CRs) indicate a strong energy dependence of the dipole amplitude and phase in the TeV--PeV range. We argue here that these observations can be well understood within standard diffusion theory as a combined effect of {\it (i)} one or more local sources at Galactic longitude $120^\circ\lesssim l\lesssim300^\circ$ dominating the CR gradient below 0.1--0.3~PeV, {\it (ii)} the presence of a strong ordered magnetic field in our local environment, {\it (iii)} the relative motion of the solar system, and {\it (iv)} the limited reconstruction capabilities of ground-based observatories. We show that an excellent candidate of the local CR source responsible for the dipole anisotropy at 1--100~TeV is the Vela supernova remnant. 
\end{abstract}


\maketitle

{\it Introduction.---}Cosmic rays (CRs) below the {\it knee} at 3--5~PeV are expected to originate in Galactic sources, presumably supernova remnants~\cite{Baade1934}. After CRs are released by their sources, they start to diffuse through the Galactic environment as a result of repeated scattering in chaotic magnetic fields. This mechanism explains the high isotropy of Galactic CR arrival directions despite the fact that their sources should align with the Galactic plane. In addition, higher energy (i.e., higher rigidity) CRs are expected to diffuse faster and leave the Milky Way more quickly. This 0dependent diffusive leakage serves as one explanation for why the locally observed CR density $n$ has a much softer spectrum, $n\propto E^{-2.7}$, than expected from diffusive shock acceleration in the sources~\cite{Bell1978a,Blandford1978}.

Standard diffusion theory predicts a small residual dipole anisotropy (DA) in the CR arrival direction. In the case of isotropic diffusion with a smooth distribution of sources, the DA is expected to simply align with the direction of the Galactic center with an amplitude following the energy scaling of the diffusion tensor, typically ${\bf K}_{\rm iso}\propto E^\beta$ with $\beta\simeq0.3$--$0.6$. Indeed, various CR, $\gamma$-ray, and neutrino observatories could identify DAs in the arrival directions of CRs at the level of $10^{-4}$--$10^{-3}$, see Ref.~\cite{DiSciascio:2014jwa} for a recent review. However, the data from recent studies of ARGO-YBJ~\cite{Bartoli:2015ysa}, EAS-TOP~\cite{Aglietta:2009mu}, IceCube/IceTop~\cite{Aartsen:2012ma,Aartsen:2016ivj}, and Tibet-AS$\gamma$~\cite{TibetICRC355} indicate that the TeV--PeV DA is not described by a simple power law and undergoes a rapid phase flip at an energy of 0.1--0.3~PeV. 

Fluctuations of the DA can be induced by the presence of local and young CR sources even if their individual contribution to the total CR flux is only subdominant~\citep{Erlykin:2006ri,Blasi:2011fi,Blasi:2011fm,Pohl:2012xs,Sveshnikova:2013ui,Kumar:2014dma}. The relative contribution of these sources compared to the overall diffuse emission from distant or old emitters depends on various aspects of the source population and the Galactic diffusion region. In general, one can expect phase rotations and amplitude modulations of the DA~\cite{Blasi:2011fm}. On the other hand, a phase {\it flip} (with vanishing amplitude) would require some fine-tuning of the contribution of local sources and the Galactic average.

For practical reasons, most scenarios discuss ensemble fluctuations of the DA under the assumption of isotropic CR diffusion~\citep{Erlykin:2006ri,Blasi:2011fi,Blasi:2011fm,Pohl:2012xs,Sveshnikova:2013ui} (but see also Ref.~\cite{Kumar:2014dma}). This ansatz seems appropriate for the prediction of the local CR density since CR diffusion from distant sources typically averages over local properties of the diffusion medium. However, the specifics of our local environment cannot be neglected when it comes to discussing the observed anisotropy. The presence of an ordered magnetic field in our local environment with a strength at the level of $3\mu$G induces circular motion of CRs around magnetic field lines with a Larmor radius $r_L\simeq 0.4 E_{\rm PeV}/(ZB_{\rm 3\mu G}) {\rm pc}$. This length scale is much smaller than the typical scattering length of TeV--PeV CRs predicted by isotropic diffusion models and indicates a strong anisotropic diffusion in our local environment.

We will show in the following that recent data on the TeV--PeV DA are consistent with the predictions of standard diffusion theory if the effects of our local environment are properly taken into account. For a correct interpretation of the CR data, it is also important to account for limited reconstruction capabilities of CR observatories often overlooked in phenomenological discussions. Ground-based CR observatories are insensitive to those CR anisotropy features that align with Earth's rotation axis. This can have a large effect on the observed dipole amplitude if the true dipole aligns with the celestial poles. In addition, the limited integrated field of view (FOV) of observatories introduces a cross talk of the dipole with higher multipole moments.

{\it Dipole anisotropy.---}The DA in the plasma rest frame (denoted by starred quantities in the following) is proportional to the spatial gradient of the CR density $\nabla n^\star$ and the diffusion tensor ${\bf K}$,
\begin{equation}\label{eq:diffusedipole}
{\boldsymbol \delta^\star} = 3{\bf K}\!\cdot\!\nabla \ln n^\star\,.
\end{equation}
In general, the diffusion tensor is expected to be invariant under rotations along the orientation of the local ordered magnetic field and can be written in the form
\begin{equation}\label{eq:K}
{K}_{ij} = \frac{\hat{B}_i\hat{B}_j}{3\nu_\parallel}+\frac{\delta_{ij}-\hat{B}_i\hat{B}_j}{3\nu_\perp}+\frac{\epsilon_{ijk}\hat{B}_k}{3\nu_A}\,.
\end{equation} 
Here, $\hat{\bf B}$ is a unit vector pointing in the direction of the regular magnetic field, $\nu_\parallel$ and $\nu_\perp$ denote the effective scattering rates along and perpendicular to the magnetic field, respectively, and $\nu_A$ is the axial scattering rate (see, e.g., Ref.~\citep{Bhatnagar:1954zz}). For TeV--PeV CRs, $\nu_\perp \gg \nu_\parallel$ and $\nu_A \gg \nu_\parallel$ and in this case the diffusion tensor (\ref{eq:K}) reduces to the first term corresponding to a projection of the CR gradient onto the magnetic field direction~\cite{JonesApJ1990,Mertsch:2014cua,Schwadron2014,Kumar:2014dma}. It was already speculated in Ref.~\cite{Mertsch:2014cua} that this projection could explain the low amplitude of the observed DA in the TeV--PeV range.

Anisotropic diffusion predicts that the DA of TeV--PeV CRs should align with the local ordered magnetic field. This ordered magnetic field corresponds to the sum of the large-scale regular magnetic field and the contribution of a chaotic component averaged over distance scales set by the Larmor radius. It has been argued that the local ordered magnetic field on distance scales less than $0.1$~pc can be inferred from the emission of energetic neutral atoms (ENA) from the outer heliosphere observed by the {\it Interstellar Boundary Explorer} (IBEX)~\cite{McComas2009}. The emission of ENA is enhanced along a circular ribbon that defines a magnetic field axis along $l \simeq 210.5^\circ$ and $b\simeq-57.1^\circ$ with an uncertainty of $\sim1.5^\circ$~\cite{Funsten2013}. Polarization measurements of local stars within $40$~pc suggest a similar field direction along $l\simeq216.2^\circ$ and $b\simeq-49.0^\circ$, although with larger statistical and systematic uncertainties~\cite{Frisch2015}.

The projection onto the regular magnetic field axis in general does not allow one to reconstruct the CR gradient, which would indicate the position of local sources. However, the phase of the projection serves as an estimate for the location of the CR gradient with respect to the magnetic hemispheres. In particular, the local magnetic field inferred from the IBEX observation divides the Galactic plane into longitude bands $120^\circ\lesssim l\lesssim300^\circ$ and the complement. The phase of the CR dipole data then allows one to allocate the CR gradient within these regions.

\begin{table}[t]\renewcommand{\arraystretch}{1.3}
\centering
\begin{ruledtabular}
\begin{tabular}{lrrrrrrr}
Observatory&$\delta_1\,\,$&$\delta_2\,\,$&$M_{11}$&$M_{12}$&$M_{13}$&$M_{14}$&$M_{15}$\\
\hline
EAS-TOP~\cite{Aglietta:2009mu}&$10^\circ$&$58^\circ$&0.828&0.891&0.262&-0.331&-0.364\\
Tibet AS-$\gamma$~\cite{TibetICRC355}&-$30^\circ$&$90^\circ$&0.842&0.323&-0.056&0.168&0.098\\
ARGO-YBJ~\cite{Bartoli:2015ysa}&-$10^\circ$&$70^\circ$&0.848&0.613&-0.015&-0.086&0.135\\
IceCube~\cite{Aartsen:2016ivj}&-$90^\circ$&-$25^\circ$&0.651&-0.961&0.789&-0.324&-0.051\\
IceTop~\cite{Aartsen:2012ma,Aartsen:2016ivj}&-$90^\circ$&-$35^\circ$&0.575&-0.961&0.999&-0.695&0.256\\
\hline
full sky&-$90^\circ$&$90^\circ$&0.785&\multicolumn{1}{c}{0}&0.184&\multicolumn{1}{c}{0}&0.091\\
\end{tabular}
\end{ruledtabular}
\caption[]{The first five mixing coefficients (\ref{eq:M1ell}) for a dipole analysis after averaging the relative CR intensity over declination. We show results for TeV--PeV observatories with an effective declination range $[\delta_1,\delta_2]$ and the result for an ideal observatory with a full FOV.}\label{tab1}
\end{table}

The relative motion of the solar system through the local plasma frame introduces an energy-independent shift of the dipole due to the Compton-Getting effect~\cite{CG1935}. The dipole in the comoving frame can be written 
\begin{equation}\label{eq:deltarel}
{\boldsymbol \delta} = {\boldsymbol \delta}^\star + (2+\Gamma)\bbeta_\odot\,,
\end{equation}
where $\bbeta_\odot = {\bf v}/c$ is the normalized velocity vector of the Sun through the plasma and $\Gamma\simeq2.7$ is the CR spectral index~\citep{Forman1970}. It should be noted that the exact rest frame of the plasma is ambiguous and should also depend on the CR rigidities under consideration. A natural choice seems to be the local standard of rest (LSR) corresponding to the Sun's motion towards the solar apex. In this case, the velocity vector points to $l \simeq 47.9^\circ\pm2.9^\circ$ and $b\simeq23.8^\circ\pm2.0^\circ$ with $v_{\rm LSR} \simeq 18.0\pm0.9$ km/s~\cite{Schoenrich2010}. We will use this estimate in the following as our nominal value. Another choice would be the relative velocity through the local interstellar medium (ISM) with orientation $l \simeq 5.25^\circ\pm0.24^\circ$ and $b\simeq12.0^\circ\pm0.5^\circ$ with $v_{\rm ISM} \simeq 23.2\pm0.3$ km/s~\cite{McComas2012}. However, we will see in the following that the relative scatter between.

{\it Observation.---} Because of the smallness of TeV--PeV CR anisotropies their experimental observation requires analysis methods that can account for uncertainties of the local detector response~\citep{Abdo:2008aw,Bonino:2011nx,Cui2003ICRC,Ahlers:2016njl}. These methods make ground-based CR observatories incapable of observing CR anisotropies along Earth's rotation axis. More precisely, if we expand the CR relative intensity into spherical harmonics $Y_{\ell m}$ in the equatorial coordinate system, all $a_{\ell0}$ coefficients are unconstrained by the observation and consequently set to $a_{\ell0}=0$ (see, e.g., Ref.~\cite{Ahlers:2016njl}). The observable dipole is then in the form 
\begin{equation}\label{eq:delta2D}
{\boldsymbol \delta}_{\rm obs} = \delta_{0{\rm h}}{\bf e}_{0{\rm h}} + \delta_{6{\rm h}}{\bf e}_{6{\rm h}}\,,
\end{equation}
where ${\bf e}_{0{\rm h}}$ and ${\bf e}_{6{\rm h}}$ are unit vectors in the equatorial plane pointing towards local sidereal time $0$h ($\alpha=0^\circ$) and $6$h ($\alpha=90^\circ$), respectively, and $\delta_{0{\rm h}} = {\boldsymbol \delta}\cdot{\bf e}_{0{\rm h}}$ and $\delta_{6{\rm h}} = {\boldsymbol \delta}\cdot{\bf e}_{6{\rm h}}$.

Because of these limited reconstruction capabilities, most CR anisotropy analyses extract an equatorial dipole amplitude $A_1$ and phase $\alpha_1$ from the declination average of the relative CR intensity $I$, 
\begin{equation}\label{eq:A}
A_1e^{i\alpha_1} = \frac{1}{\pi(s_2-s_1)}\int_0^{2\pi}{\rm d}\alpha\int_{\delta_1}^{\delta_2}{\rm d}\delta\cos\delta e^{i\alpha}I(\alpha,\delta)\,,
\end{equation}
where $s_{1/2}=\sin\delta_{1/2}$ and $[\delta_1,\delta_2]$ is the declination interval of the observatories' time-integrated FOV (see Table~\ref{tab1}). Note that expression (\ref{eq:A}) assumes that any intensity variation induced by the local detector acceptance is corrected, e.g., by following the method described in Ref.~\citep{Ahlers:2016njl}.

Recent observations indicate that there exist significant anisotropies in the arrival direction of CRs down to angular scales of $10^\circ$~\cite{Abeysekara:2014sna,Aartsen:2016ivj}. Hence, the observed relative CR intensity must be expressed as a sum over spherical harmonics in the form
\begin{equation}\label{eq:I}
I(\alpha,\delta)=1+\sum_{\ell\geq1}\sum_{m\neq0}a_{\ell m}Y_{\ell m}(\alpha,\pi/2-\delta)\,,
\end{equation}
where $a_{\ell m}$ are complex numbers obeying $a_{\ell -m} = (-1)^m a^*_{\ell m}$. The DA is expressed in terms of the $\ell=1$ coefficient as $a_{1-1} = (\delta_{0{\rm h}} + i\delta_{6{\rm h}})\sqrt{2\pi/3}$. With the most general relative intensity (\ref{eq:I}), the integral (\ref{eq:A}) can be rewritten in the form $A_1e^{i\alpha_1} = \sqrt{3/2\pi}\sum_\ell M_{1\ell}a_{\ell -1}$, with 
\begin{equation}\label{eq:M1ell}
M_{1\ell} = \frac{1}{s_1-s_2}\sqrt{\frac{2(2\ell+1)}{3\ell(\ell+1)}}\int_{s_1}^{s_2}{\rm d}s P_\ell^1(s)\,,
\end{equation}
where $P_\ell^1$ are the associated Legendre polynomials and $s_{1/2}=\sin\delta_{1/2}$. 

Table~\ref{tab1} shows the mixing terms $M_{1\ell}$ for the first five multipole moments for five different CR observatories. The mixing terms are large and indicate that the presence of medium-scale anisotropy can have a significant effect on the DA. We also show the mixing of multipole moments derived from the one-dimensional dipole analysis for an ideal observatory with full sky coverage. In the case of a large integrated FOV, the dipole analysis in terms of a two-dimensional analysis of spherical harmonics seems more appropriate (see, e.g., Ref.~\cite{Ahlers:2016njl}).

\begin{figure}[t]
\centering
\includegraphics[width=0.95\linewidth]{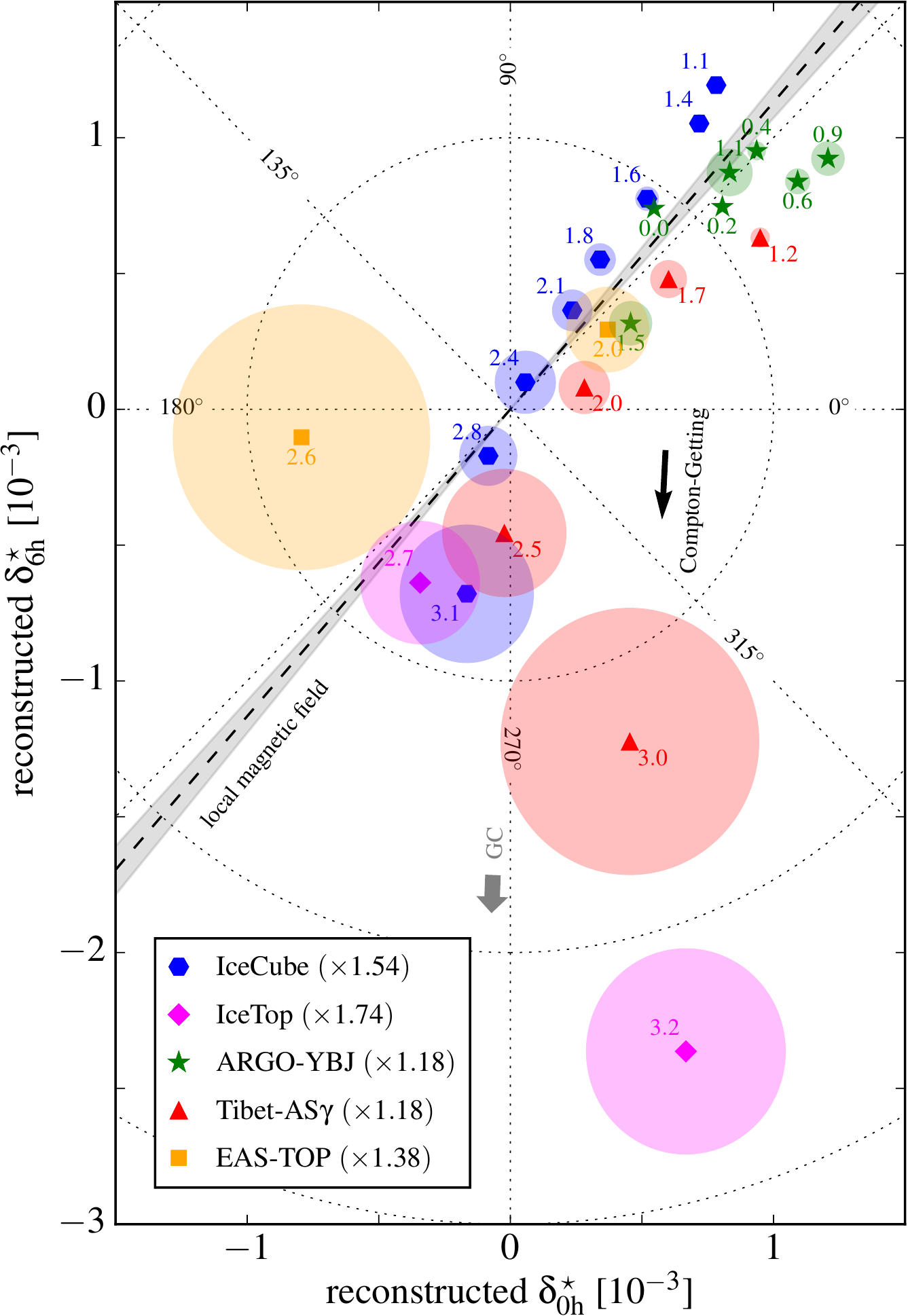}
\caption[]{Summary plot of the reconstructed TeV--PeV dipole components $\delta^\star_{0{\rm h}}$ and $\delta^\star_{6{\rm h}}$ in the equatorial plane. The black arrow indicates the Compton-Getting effect from the solar motion with respect to the local standard of rest that we subtracted from the data following Eq.~(\ref{eq:deltarel}). We follow Eq.~(\ref{eq:conversion}) to rescale the declination-averaged data of ARGO-YBJ~\cite{Bartoli:2015ysa}, Tibet-AS$\gamma$~\cite{TibetICRC355}, and IceCube/IceTop~\cite{Aartsen:2012ma,Aartsen:2016ivj}. We also include results by EAS-TOP~\cite{Aglietta:2009mu} derived by the East-West method (see Supplemental Material). The numbers attached to the data indicate the median energy of the bins as $\log_{10}(E_{\rm med}/{\rm TeV})$. The colored disks show the $1\sigma$ error range estimated by Eq.~(\ref{eq:deltadelta}). The dashed line and gray-shaded area indicate the magnetic field direction and its uncertainty (projected onto the equatorial plane) inferred from IBEX observations~\cite{Funsten2013}. We also indicate the direction towards the Galactic center (GC).}\label{fig1}
\end{figure}

However, under the assumption that higher multipole moments are negligible, $|a_{\ell -1}|\ll |a_{1 -1}|$ for $\ell > 1$, we can proceed by estimating the projected dipole component as
\begin{equation}\label{eq:conversion}
({\delta}_{0{\rm h}},{\delta}_{6{\rm h}}) \simeq \frac{1}{M_{11}}(A_1\cos\alpha_1,A_1\sin\alpha_1)\,.
\end{equation}
The statistical uncertainty of the dipole vector can be expressed via the uncertainties on the amplitude and phase as
\begin{equation}\label{eq:deltadelta}
|\Delta({\delta}_{0{\rm h}},{\delta}_{6{\rm h}})|^2 \simeq \frac{(\Delta A_1)^2+A_1^2(\Delta\alpha_1)^2}{M^2_{11}}\,.
\end{equation}
Figure~\ref{fig1} shows a summary of TeV--PeV anisotropy measurements from the observatories listed in Table~\ref{tab1}. In order to convert the amplitude and phase from declination-averaged data to the true horizontal dipole components we use the conversion (\ref{eq:conversion}) and correct for a Compton-Getting effect of the solar motion in the LSR in the relation (\ref{eq:deltarel}). The EAS-TOP dipole measurement~\cite{Aglietta:2009mu} is inferred by the East-West method and in this case the relation between measured and true dipole follows a different relation (see Supplemental Material). The individual data sets presented in Fig.~\ref{fig1} show a large relative scatter at similar median energies, even after correcting for the partial sky coverage of the observatories. This indicates the contribution of cross talk of the dipole with $\ell\gtrsim2$ multipoles of the individual observatories (see Table~\ref{tab1}).

The dashed line in Fig.~\ref{fig1} shows the best-fit magnetic field direction inferred from the IBEX observation~\cite{Funsten2013} with uncertainties indicated by gray-shaded wedges. The analysis of Ref.~\cite{Schwadron2014} already pointed out that the combined anisotropy maps of IceCube at 20~TeV and Tibet-AS$\gamma$ at 5~TeV show a close alignment of large-scale features with this local magnetic field direction as a result of anisotropic diffusion. One can see that this interpretation is consistent with the general trend of the TeV--PeV data collected by other observatories. Note that the phase of the dipole below 0.1--0.3~PeV indicates that the CR gradient aligns with Galactic longitudes $120^\circ\lesssim l\lesssim300^\circ$. Hence, the CR anisotropy below this energy is expected to follow the contribution of one or more local sources, rather than the average CR gradient pointing towards the Galactic center region.

The CR anisotropy data beyond 1~PeV are inconclusive. The IceTop and Tibet-AS$\gamma$ data suggest a trend away from the IBEX fit. However, recent measurements of the PeV CR anisotropy by KASCADE-Grande~\cite{KASCADEICRC281} show dipole phases of $\alpha(3{\rm PeV})\simeq225^\circ\pm 22^\circ$ and $\alpha(6{\rm PeV})\simeq227^\circ\pm 30^\circ$, consistent with the IBEX observation. Because of the low significance of the corresponding amplitudes ($3.7\sigma$) these data points were not shown in Fig.~\ref{fig1}. Note that the Larmor radius of PeV CR energies approaches $r_L\simeq 0.4$~pc which extends to outside the edge of the local cloud~\cite{Frisch2011}. It is therefore feasible that the average ordered magnetic field changes orientation at this energy.

{\it A plausible scenario: The Vela SNR.---}Before we conclude, let us consider one example showing the contribution of a local source on the CR anisotropy in the presence of a strong local magnetic field. For our calculation, we assume that the spectrum of CRs from individual sources can be derived from an effective isotropic diffusion tensor with $K_{\rm iso} \simeq 4\times10^{28}(E/3{\rm GeV})^{1/3} {\rm cm}^2/{\rm s}$ and half height $H=3$~kpc of the diffusion region and a Galactic SNR rate of $\mathcal{R}_{\rm SNR}=1/30~{\rm yr}^{-1}$ (see Supplemental Material). The sources are assumed to follow an azimuthally symmetric distribution~\cite{Case:1998qg}. The local magnetic field in our example aligns with the direction inferred by IBEX and acts as a projection operator, assuming that the large-scale effective isotropic diffusion rate coincides with the local parallel diffusion rate.

The gradient of a local source at distance $d$ and emission time $\tau$ reaches a maximum at an energy satisfying $6K_{\rm iso}\tau=d^2$ with a contribution scaling as $3K_{\rm iso}|\nabla n_{\star}|\simeq 0.11Q_{\star}/(d^2c\tau)$, where $Q_{\star}$ is the time-integrated CR spectrum. Among the list of known local SNRs~\cite{Green:2014cea} Vela ($l = 263.9^\circ$, $b=-3.3^\circ$), at a distance of about $0.3$~kpc~\cite{Cha:1999pn}, with an age of $11$~kyr~\cite{Reichley1970}, and with an ejecta energy of $10^{51}$~erg~\cite{Jenkins1995} is expected to have the strongest local contribution to the CR anisotropy. The analysis of Ref.~\cite{Sveshnikova:2013ui} already recognized the importance of the Vela SNR for the CR anisotropy in the TeV region but did not account for magnetic projection effects.

\begin{figure}[t]
\centering
\includegraphics[width=0.95\linewidth]{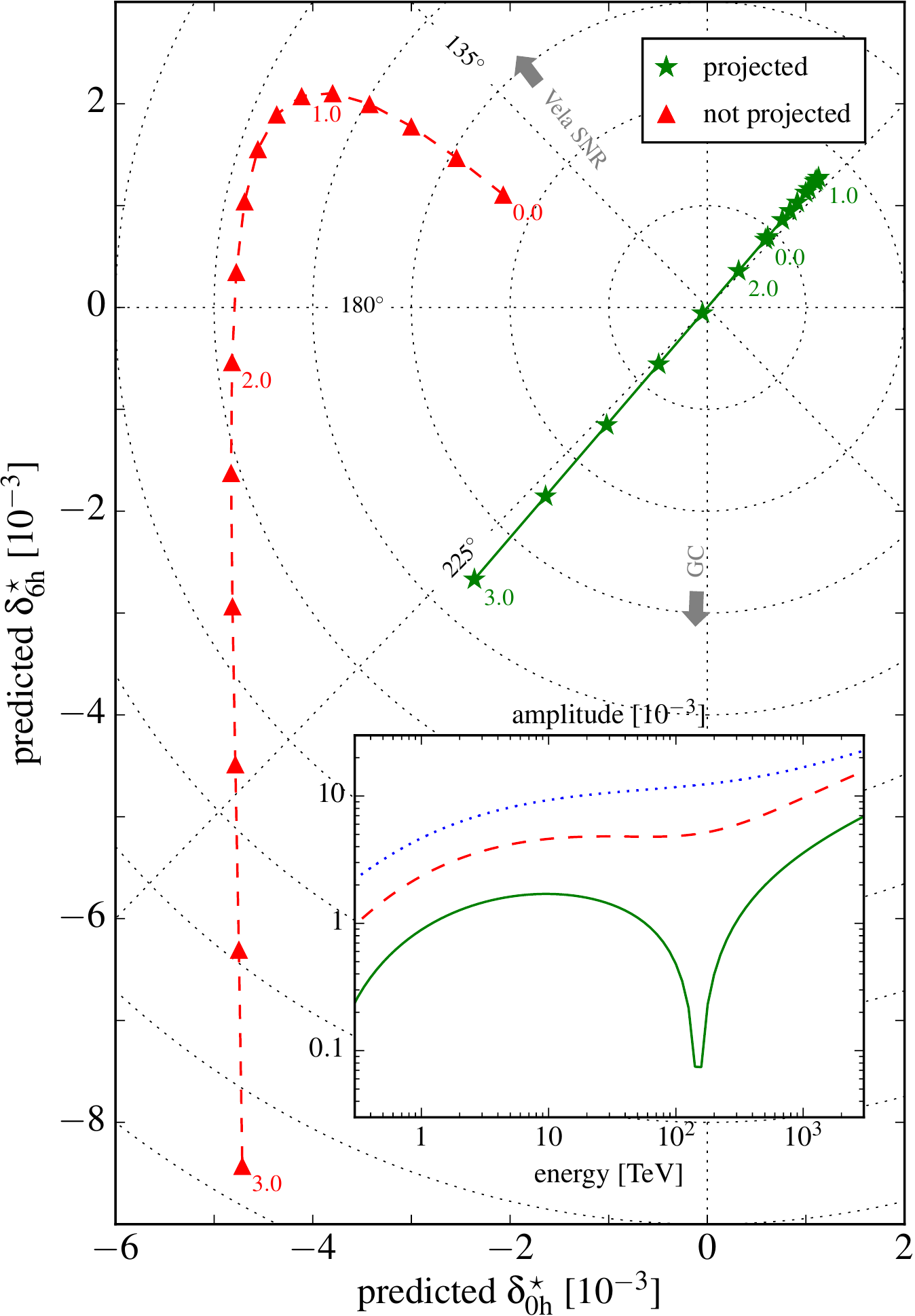}
\caption[]{An example showing the effect of the Vela SNR on the average anisotropy from all Galactic SNRs. The data points show the projected TeV--PeV DA in equal logarithmic energy steps. The red triangles and green stars shows the anisotropy evolution with and without magnetic field projections, respectively. The inset plot shows the amplitude of the two simulations. The blue dotted line shows the naive expectation of the full three-dimensional dipole amplitude without projection onto the magnetic field and into the equatorial plane.}\label{fig2}
\end{figure}

Assuming instantaneous emission of CRs at the beginning of the Sedov phase after 100 years, the maximum gradient is reached for an energy around a few TeV with a density of $3K_{\rm iso}|\nabla n_{\star}|\simeq 0.34 Q_{\star}/{\rm kpc}^3$ and is then expected to quickly fall off. On the other hand, the cumulative anisotropy from all Galactic sources is expected to be flat in this energy range and should point towards the Galactic center region. Numerically, we find that the average contribution of all SNRs has a gradient $3K_{\rm iso}|\nabla \langle n\rangle| \simeq 0.11\langle Q_{\star}\rangle /{\rm kpc}^3$. Therefore, in this setup, the Vela SNR can dominate the anisotropy around 10~TeV and its position falls within $120^\circ\lesssim l\lesssim300^\circ$, consistent with the observed phase of the DA.

The data points shown in Fig.~\ref{fig2} show the predicted DA from the combined contribution of Vela and the Galactic average projected onto the equatorial plane. The green stars and red triangles show the expected anisotropy with and without the magnetic projection, respectively, in steps of $\log_{10} (E/{\rm TeV})= 0.2$. The projected data is qualitatively very similar to the TeV--PeV data shown in Fig.~\ref{fig1}. The inset plot shows the amplitude of the observation with (solid green) and without (red dashed) magnetic projection. The blue dotted line shows also the full predicted dipole amplitude, which is not observable by ground-based detectors.

{\it Conclusions.---}In this study we have shown that the observed CR anisotropy in the TeV-PeV is consistent with the paradigm of CR diffusion in the Galactic environment if one takes into account the effect of local magnetic fields and (to a lesser extent) the relative motion of the solar system. For the interpretation of CR data, it is important to account for large systematic uncertainties of CR observation. Let us conclude with a few final remarks.

{\it (i)} The traditional analysis method of averaging the observed relative intensity over declination introduces cross talk between multipole moments even in the ideal case of a full sky coverage. In this respect, a two-dimensional analysis seems more appropriate. The remaining cross talk of multipole moments from the limited FOV (weight function) can be reduced by a joint analysis of CR data~\cite{Ahlers:2016njl,TibetICRC279,HAWCICRC444}. 

{\it (ii)} The expected DA is a projection of the local CR gradient onto the magnetic field direction. The observed DA that results from the projection onto the Celestial equator is, hence, a measure of the local ordered magnetic field averaged over distance scales corresponding to the CR Larmor radius. Depending on the size of systematic uncertainties of the CR data, this can serve as a new measure of the local magnetic field direction.

{\it (iii)} The projection of the TeV CR dipole onto the magnetic field axis does not allow one to reconstruct the CR gradient. However, we can determine the magnetic hemisphere of the CR gradient by the phase. For the best-fit local magnetic field inferred by the IBEX observation, the alignment of the TeV dipole indicates a source with $120^\circ\lesssim l\lesssim300^\circ$. We have shown that Vela SNR could be responsible for the CR gradient in this hemisphere with a transition to a gradient at $l\simeq0^\circ$ from the average CR source distribution at higher energies.

\medskip

M.~A.~thanks Feng Zhaoyang and Kauoki Munakata for comments regarding the analyses of Refs.~\cite{TibetICRC279,TibetICRC355} as well as Francis Halzen and Stefan Westerhoff for comments on the manuscript. This work is supported by the National Science Foundation (Grants No.~PHY-1306958 and No.~PLR-1600823).

\appendix

\section{Supplemental Material}

\subsection{East-West Method}

The dipole data by EAS-TOP~\cite{Aglietta:2009mu} shown in Fig.~\ref{fig1} is reconstructed via the East-West method. This method also estimates the dipole amplitude and phase of the relative intensity after averaging the data over declination. However, in contrast to expression (\ref{eq:A}), the dipole is derived from the {\it derivative} of the relative intensity with respect to right ascension.

Let's assume that the relative intensity is given by a dipole, $I(\alpha,\delta) = 1 + \boldsymbol{\delta}\!\cdot\!{\bf n}(\alpha,\delta)$, where ${\bf n}$ is a unit vector in the equatorial coordinate system. It is related to the corresponding unit vector ${\bf n}'(\varphi,\theta)$ in the local coordinate system with azimuth angle $\varphi$ and zenith angle $\theta$ via a time-dependent rotation matrix, ${\bf n}'={\bf R}(t)\cdot{\bf n}$. (Explicit expressions of ${\bf n}$, ${\bf n}'$, and ${\bf R}$ can be found, e.g., in Ref.~\cite{Ahlers:2016njl}.)

Now, at each sidereal time $t$ the CR data is divided into two bins, covering mirror-symmetric portions of the East ($0<\varphi<\pi$) and West ($-\pi<\varphi<0$) sectors in the local coordinate system. The event numbers observed during a short sidereal time interval $\Delta t$ are then related to the relative intensity $I$ and total accumulated detector exposure $\mathcal{E}$ as
\begin{align}
N_{\rm E}(t) &\simeq \frac{nc\Delta t }{4\pi}\int_{\varphi_1}^{\varphi_2}{\rm d}\varphi\!\!\!\int_0^{\theta_{\rm max}}\!\!{\rm d}\theta\sin\theta\,\mathcal{E}(t,\varphi,\theta)I(t,\varphi,\theta)\,,\\
N_{\rm W}(t) &\simeq \frac{nc\Delta t}{4\pi}\!\!\int_{-\varphi_2}^{-\varphi_1}\!\!{\rm d}\varphi\!\!\!\int_0^{\theta_{\rm max}}\!{\rm d}\theta\sin\theta\,\mathcal{E}(t,\varphi,\theta)I(t,\varphi,\theta)\,,\end{align}
with $0<\varphi_1<\varphi_2<\pi$.
Now, a crucial assumption of this method is that the exposure $\mathcal{E}$ can be expressed as a product of its angular-integrated exposure $E$ and relative acceptance $\mathcal{A}$ depending only on zenith angle $\theta$,
\begin{equation}\label{eq:E}
  \mathcal{E}(t,\varphi,\theta) \simeq E(t)\mathcal{A}(\theta)\,.
\end{equation}
Then, the relative difference between the East and West data is independent of the absolute exposure $E(t)$, and we arrive at
\begin{equation}
\frac{N_{\rm E}(t)-N_{\rm W}(t)}{N_{\rm E}(t)+N_{\rm W}(t)} \simeq \Delta\alpha\frac{\partial}{\partial\alpha}\delta I(\alpha,0)\,,
\end{equation}
where $\Delta\alpha$ is an effective right ascension step size. For the dipole anisotropy, it can be calculated as
\begin{equation}\label{eq:dalpha}
\Delta\alpha = \langle\sin\theta\sin\varphi\rangle_\mathcal{A}\,,
\end{equation}
where $\langle\cdot\rangle_\mathcal{A}$ indicates the average over the East (or West) sector with weight $\mathcal{A}(\theta)$.

The EAS-TOP analysis~\cite{Aglietta:2009mu} uses East/West bins with azimuthal range limited by $\varphi_1=45^\circ$ and $\varphi_2=135^\circ$, and zenith angle cut $\theta_{\rm max}=40^\circ$. The effective right ascension step in expression (\ref{eq:dalpha}) is approximated by the average hour angle measured from zenith, $\Delta\alpha \simeq \langle \delta t\rangle_\mathcal{A}$. The zenith angle distribution of CRs at the position of EAS-TOP is approximately $\mathcal{A} \propto \exp(-n/\cos\theta)$ with $n\simeq 6.6$~\cite{Aglietta:1993gq}. Numerically, we can reproduce $\langle\delta t\rangle_\mathcal{A} \simeq 1.71$h ($1{\rm h} = 15^\circ$), close to the value of $1.7$h quoted by EAS-TOP~\cite{Aglietta:2009mu}. However, the exact expression (\ref{eq:dalpha}) for the same effective area gives $\Delta\alpha \simeq 1.24$~h. This corresponds to a correction factor of $1.71/1.24\simeq1.38$, which we use in Fig.~\ref{fig1} to rescale the EAS-TOP data.

Note that the East-West method also introduces cross talk between small- and large-scale multipole moments. The relative coupling is expected to follow Eq.~(7) for an effective declination range of $\delta_1=10^\circ$ and $\delta_2=58^\circ$, corresponding to the time-integrated field of view of the East (or West) sector (see Table~\ref{tab1}).

\begin{table}[t]
\renewcommand{\arraystretch}{1.3}
\begin{ruledtabular}
\begin{tabular}{lccccc}
SNR&$l$&$b$&$d$~[pc] &$T_{\rm age}$~[kyr]&Refs.\\
\hline
Loop I&$329.0^\circ$&$17.5^\circ$&$170$&$200$&\cite{Berkhuijsen1971,Egger1995}\\
Vela&$263.9^\circ$&$-3.3^\circ$&$300$&$11$&\cite{Reichley1970,Cha:1999pn}\\
Monogem&$203.0^\circ$&$12.0^\circ$&$300$&$86$&\cite{Plucinsky1996}\\
Geminga&$197.0^\circ$&$-11.7^\circ$&$400$&$340$&\cite{Gehrels1993,Caraveo1996}\\
Cygnus Loop&$74.0^\circ$&$-8.6^\circ$&$440$&$20$&\cite{Levenson1997,Blair1999}
\end{tabular}
\end{ruledtabular}
\caption[]{The position and age of the five closest known SNRs.}\label{tab2}
\end{table}

\subsection{Local Cosmic Ray Density}

In the following, we discuss solutions to the isotropic diffusion equation given by
\begin{equation}\label{eq:diffusion}
\partial_tn - K\Delta n = \mathcal{Q}\,.
\end{equation}
We will work with galactocentric coordinates, where the position of the solar system is ${\bf r}_\odot = (-R_\odot,0,0)$. The position of a source at distance $d$ and observed at Galactic longitude $l$ and latitude $b$ is then given by $x_s=d\cos l\cos b-R_\odot$, $y_s=d\sin l\cos b$,  and $z_s=d\sin b$.

Here, we consider the case that CR diffusion is limited to a region bounded by $|z|\leq H$ and the CR density vanishes on the boundary, $n(z=\pm H)=0$. The appropriate Green function of this problem can then be derived from the free Green function of the source term $\mathcal{Q} = \delta^{(3)}({\bf r}-{\bf r}_s)\delta(t-t_s)$ with the technique of mirror charges (see, e.g., Refs.~\cite{Blasi:2011fi,Blasi:2011fm}). For $t>t_s$ we have
\begin{equation}\label{eq:Green}
G({\bf r},t,{\bf r}_s,t_s) = (4\pi K\tau_s)^{-\frac{3}{2}}\sum_{n=-\infty}^\infty (-1)^ne^{-\tfrac{({\bf r}-{\bf r}_n)^2}{4K\tau_s}}\,,
\end{equation}
with ${\bf r}_n \equiv (x_s,y_s,(-1)^nz_s+2nH)$ and $\tau_s \equiv t-t_s$. We approximate the time-scale $\tau_s$ that has passed since the (instantaneous) emission of CRs as $\tau_s=T_{\rm age}-T_{\rm Sedov}+d/c$, where $T_{\rm Sedov} \simeq 100$~yr marks the beginning of the Sedov phase and $d/c$ accounts for the light-travel time.

\begin{figure}[t]
\centering
\includegraphics[width=0.95\linewidth]{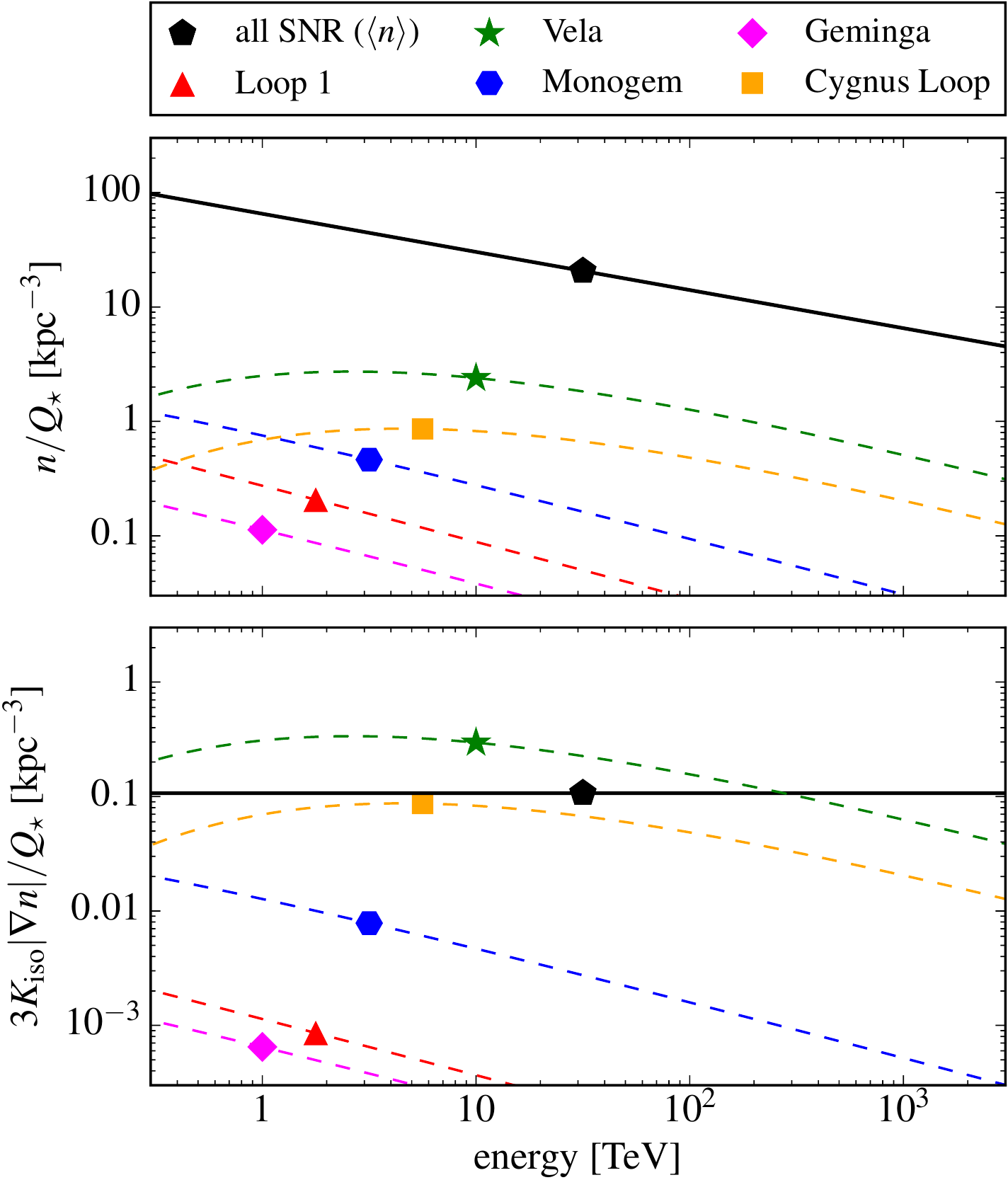}
\caption[]{The contribution of the five closest SNRs to the CR density (top) and CR gradient (bottom).}\label{fig3}
\end{figure}

The Green function (\ref{eq:Green}) describes the contribution of a CR point source that instantaneously emits CRs with a spectrum $Q_\star(E)$,
\begin{equation}\label{eq:PS}
n_\star = Q_\star G({\bf r}_\odot,0,{\bf r}_s,-\tau)\,.
\end{equation}
The ensemble-averaged contribution of {\it all} Galactic sources can be calculated from the appropriate source distribution. Here, we use the distribution of supernova remnants (SNR) following Ref.~\cite{Case:1998qg}. The probability distribution can be expressed in terms of galactocentric cylindrical coordinates as 
\begin{equation}\label{eq:Case}
\rho(r,z) = \rho_0\left(\frac{r}{R_\odot}\right)^\alpha e^{-\beta\tfrac{r-R_\odot}{R_\odot}}e^{-\tfrac{|z|}{h}} \,,
\end{equation}
with $\alpha=2$, $\beta=3.53$, $R_\odot=8.5$~kpc and $h=0.181$~kpc. The normalization $\rho_0$ is here chosen such that $2\pi\int{\rm d}r r \int{\rm d}z\rho(r,z) = 1$. The ensemble-averaged contribution is then given by
\begin{equation}\label{eq:smooth}
\langle n\rangle= \langle{Q}_\star\rangle\mathcal{R}_{\rm SNR}\int_0^\infty{\rm d}r r\!\int_{-H}^H{\rm d}z\int_0^{2\pi}{\rm d}\alpha\!\!\sum_{n=-\infty}^\infty\frac{(-1)^n\rho(r,z)}{4\pi K|{\bf r}_n-{\bf r}_\odot|}\,,
\end{equation}
with ${\bf r}_{\rm n} = (r\cos\alpha,r\sin\alpha,(-1)^nz+2nH)$. The factor $\mathcal{R}_{\rm SNR}$ is the Galactic supernova rate and $\langle Q_\star\rangle$ denotes the ensemble-averaged emission spectrum of the sources.

We can now estimate the effect of a single local source by the sum $n \simeq {n}_\star + \langle{n}\rangle$ and the corresponding equation for the gradient. Note that in the case of universal source spectra, $Q_\star =\langle Q_\star\rangle$, the dipole (\ref{eq:diffusedipole}) is independent of the source spectrum. The energy dependence is then entirely determined by the dependence of the Green function on the diffusion coefficient.

In Figure~\ref{fig3} we show the density (top panel) and density gradient (lower panel) of the smooth distribution (\ref{eq:smooth}) (solid lines) and for the five closest known SNRs (dashed lines) listed in Table~\ref{tab2}. The calculation assumes the same parameters used in the main text, i.e., a source rate $\mathcal{R}_{\rm SNR} = 1/30\,{\rm yr}^{-1}$, a vertical diffusion height of $H=3$~kpc, and an (effective) isotropic diffusion coefficient $K \simeq 4\times10^{28}(E/3{\rm GeV})^{1/3} {\rm cm}^2/{\rm s}$.

Whereas the contribution of the local sources to the overall density is negligible, the gradient of Vela can dominate the CR gradient below 100~TeV. Another strong contribution is expected from the Cygnus Loop, in particular if the emission were slightly larger than the average. However, the Cygnus Loop appears in a different magnetic hemisphere ($l\simeq74^\circ$) and its dominance would be inconsistent with the observed dipole data.

\end{document}